# JASMIN: Japanese-American Study of Muon Interactions and Neutron Detection*†

**Hiroshi Nakashima[1], Nikolai V. Mokhov[2], Yoshimi Kasugai[1], Norihiro Matsuda[1], Yosuke Iwamoto[1], Yukio Sakamoto[1], Anthony F. Leveling[2], David J. Boehnlein[2], Kamran Vaziri[2], Toshiya Sanami[3], Hiroshi Matsumura[3], Masayuki Hagiwara[3], Hiroshi Iwase[3], Syuichi Ban[3], Hideo Hirayama[3], Takashi Nakamura[4], Koji Oishi[5], Nobuhiro Shigyo[6], Hiroyuki Arakawa[6], Tsuyoshi Kajimoto[6], Kenji Ishibashi[6], Hiroshi Yashima[7], Shun Sekimoto[7], Norikazu Kinoshita[8], Hee-Seock Lee[9] and Koji Niita[10]**

[1] Japan Atomic Energy Agency, Tokai, Ibaraki, 319-1195 Japan
[2] Fermi National Accelerator Laboratory, Batavia, IL 60510-5011 USA
[3] High Energy Accelerator Research Organization, Oho, Tsukuba, 305-0801 Japan
[4] Tohoku University, Aoba, Sendai, 980-8578 Japan
[5] Shimizu Corporation, Etchujima, Koto, Tokyo, 135-8530 Japan
[6] Kyushu University, Motooka, Fukuoka, 819-0395 Japan
[7] Kyoto University, Kumatori, Osaka 590-0494 Japan
[8] Tsukuba University, Tsukuba, 305-8577 Japan
[9] Pohang Accelerator Laboratory, POSTECH, Pohang, Kyungbuk 790-784, Korea
[10] Research Organization for Information Science & Technology, Tokai, Ibaraki, 319-1106

---

*Work supported by grant-aid of ministry of education (KAKENHI 19360432 and 21360473) in Japan, and by a U.S. Department of Energy Laboratory operated under Contract DE-AC02-07CH11359 by the Fermi Research Alliance, LLC.
†Presented at *10th Workshop on Shielding Aspects of Accelerators, Targets and Irradiation Facilities (SATIF-10)*, June 2-4, 2010, CERN, Geneva, Switzerland.

# JASMIN: Japanese-American Study of Muon Interactions and Neutron detection

**Hiroshi Nakashima[1], Nikolai V. Mokhov[2], Yoshimi Kasugai[1], Norihiro Matsuda[1], Yosuke Iwamoto[1], Yukio Sakamoto[1], Anthony F. Leveling[2], David J. Boehnlein[2], Kamran Vaziri[2], Toshiya Sanami[3], Hiroshi Matsumura[3], Masayuki Hagiwara[3], Hiroshi Iwase[3], Syuichi Ban[3], Hideo Hirayama[3], Takashi Nakamura[4], Koji Oishi[5], Nobuhiro Shigyo[6], Hiroyuki Arakawa[6], Tsuyoshi Kajimoto[6], Kenji Ishibashi[6], Hiroshi Yashima[7], Shun Sekimoto[7], Norikazu Kinoshita[8], Hee-Seock Lee[9] and Koji Niita[10]**

[1] Japan Atomic Energy Agency, Tokai, Ibaraki, 319-1195 Japan
[2] Fermi National Accelerator Laboratory, Batavia, IL 60510-5011 USA
[3] High Energy Accelerator Research Organization, Oho, Tsukuba, 305-0801 Japan
[4] Tohoku University, Aoba, Sendai, 980-8578 Japan
[5] Shimizu Corporation, Etchujima, Koto, Tokyo, 135-8530 Japan
[6] Kyushu University, Motooka, Fukuoka, 819-0395 Japan
[7] Kyoto University, Kumatori, Osaka 590-0494 Japan
[8] Tsukuba University, Tsukuba, 305-8577 Japan
[9] Pohang Accelerator Laboratory, POSTECH, Pohang, Kyungbuk 790-784, Korea
[10] Research Organization for Information Science & Technology, Tokai, Ibaraki, 319-1106 Japan

**Abstract**

*Experimental studies of shielding and radiation effects at Fermi National Accelerator Laboratory (FNAL) have been carried out under collaboration between FNAL and Japan, aiming at benchmarking of simulation codes and study of irradiation effects for upgrade and design of new high-energy accelerator facilities. The purposes of this collaboration are (1) acquisition of shielding data in a proton beam energy domain above 100GeV; (2) further evaluation of predictive accuracy of the PHITS and MARS codes; (3) modification of physics models and data in these codes if needed; (4) establishment of irradiation field for radiation effect tests; and (5) development of a code module for improved description of radiation effects. A series of experiments has been performed at the Pbar target station and NuMI facility, using irradiation of targets with 120 GeV protons for antiproton and neutrino production, as well as the M-test beam line (M-test) for measuring nuclear data and detector responses. Various nuclear and shielding data have been measured by activation methods with chemical separation techniques as well as by other detectors such as a Bonner ball counter. Analyses with the experimental data are in progress for benchmarking the PHITS and MARS15 codes. In this presentation recent activities and results are reviewed.*

## 1. Background

Since SATIF3, an international comparison has been carried out as one of the activities of SATIF, in order to compare the calculation results from an aspect of accelerator shielding. In the activity, neutron attenuation in iron and concrete has been compared, and it is pointed out since SATIF7 that some discrepancies among the codes have been observed in the energy region above a few tens of GeV, especially more than 100 GeV, as shown in Fig. 1 [1-2]. It is essential to compare with measurements to understand possible reasons for these discrepancies. The shielding experiment at CERF [3] gave already good information for benchmarking the codes for a 120-GeV beam comprised of protons and pions. At the same time, it became clear that further well-conditioned benchmark experiments are still needed for beam energies above 100 GeV.

Several high-intensity high-energy accelerator facilities such as EURISOL and Project-X of FNAL are planned in the world [4-5]. It is critical to estimate radiation damage to the structural materials for such accelerator facilities, with Displacement-per-Atom (DPA) in materials being one of the key parameters for a facility design. Some codes have the capability of calculating DPA for estimating radiation effects on materials. The accuracy of the estimations of radiation effects, however, has not yet comprehensively been evaluated by experiments in realistic radiation fields.

At SATIF7, an experiment on shielding and radiation effects using accelerator complex at FNAL has been proposed by Dr. N. Mokhov, and some budget has been set aside by Japanese colleagues. Thus, JASMIN: Japanese-American Study of Muon Interactions and Neutron detection, has been started under collaboration among many institutes and universities of Japan and U.S.A. In this paper, the present status of JASMIN on shielding experiment is reviewed [6-17].

## 2. JASMIN Goals

The purposes of JASMIN are as follows:

(1) acquiring shielding data in a proton beam energy region above 100 GeV;

(2) further evaluation of predictive accuracy of the PHITS and MARS codes [18-19];

(3) modification of physics models and data in these codes if needed;

(4) characterization of radiation fields for studies of radiation effects;

(5) development of a code module for an improved description of radiation effects.

This is to improve the confidence in applications of the codes to shielding design of high-energy accelerator facilities as well as for various research fields such as high-energy physics, space, biological and medical sciences, and nuclear and material engineering.

## 3. Experiments and Results

In this study, the following three measurements are carried out since the first campaign of FY 2007:

(1) Measurement of particle fluxes and spectra as well as residual activity around shield at the Pbar station with a 120-GeV proton beam.

(2) Measurement of particle fluxes and spectra as well a residual activity around shield at NuMI with 120-GeV protons.

(3) Thick target yield measurements in the Meson area with 120-GeV protons.

In this paper the first two measurements are reviewed, and the last one is reported in detail in other paper by Dr. T. Sanami in SATIF10 [20].

### *3.1 Measurement of particle flux, spectra and residual activity around shield at the Pbar station with 120GeV proton*

Figure 2 shows a cross sectional view of the Pbar target station. An antiproton production target, consisted of Inconel 600 disks covered by Berylium, is irradiated by 120-GeV protons with a 1.6-μs pulse width and a 2.2-sec repetition period. The typical number of protons in the pulse was $7\times10^{12}$, and the resultant beam power corresponded to about 61 kW at the first campaign. A collection lithium lens, collimator and a pulsed magnet are placed at the downstream of the target to focus, collimate and extract the produced antiprotons. The remaining proton and secondary particles are absorbed in a dump placed after the pulsed magnet. The dump consists of a graphite cylinder 20 cm in diameter and 120-cm long encapsulated in a 20-cm thick aluminum shell. Shields made of iron and concrete are placed on the upper side of the target and magnets. The thickness of iron and concrete above the target are 183 and 122 cm, respectively. A 183-cm air gap is between the iron and concrete shields [8].

In this experiment the following measurements have been planned and carried out:

(1) Measurement of neutron flux inside and outside the iron and concrete shields.

(2) Measurement of activities around the Pbar target.

(3) Measurement of air and water activation in vault.

(4) Measurement of streaming particle passing through duct from vault to outside.

In the measurements, various techniques were used: activation method, activation method with radiochemical separation, Bonner sphere counters (current mode, pulse mode), NE213 scintillation counter, phoswitch detector. Table 1 summarizes the detector parameters outside the concrete shield. The sizes of the activation detectors were changed depending on the flux intensity at each position for each measurement. Because of the high intensity of the secondary particles (mainly neutrons) at the top of the target station, it was very difficult to use some counters there.

Figure 3 shows neutron reaction rate distributions of various reactions inside the iron shield in vertical direction to proton beam axis as a typical experimental data. In order to obtain the reaction rates in the wide energy region up to about 100 MeV, reactions of In, Al, Nb and Bi having various threshold energies were selected: $^{27}$Al(n, α)$^{24}$Na (threshold energy $E_{th}$ about 3.3 MeV), $^{93}$Nb(n, 2n)$^{92m}$Nb ($E_{th}$ ~ 9.1 MeV), $^{209}$Bi(n, 4n)$^{206}$Bi ($E_{th}$ ~ 22.6 MeV), $^{209}$Bi(n, 5n)$^{205}$Bi ($E_{th}$ ~ 29.6 MeV), $^{209}$Bi(n, 6n)$^{204}$Bi ($E_{th}$ ~ 38 MeV) and $^{209}$Bi(n, 7n)$^{203}$Bi ($E_{th}$ ~ 45.3 MeV). After irradiation, radioactivity produced in each detector was measured by the HP-Ge detectors. Experimental errors range from ± 5% to several tens of % depending on the activation counting statistics. A parameter of the Moyer model [21], attenuation length λ, was deduced for the measured attenuation of each reaction rate by using the least square fitting with the Moyer model, as summarized in the figure. The average attenuation length is 150± 5 g/cm$^2$. This is the first data for incident proton energy above 100 GeV [14].

The deduced attenuation length is compared with the previous data measured at other accelerator facilities such as KEK and CERN, as a function of maximum source neutron energy, as shown in Fig. 4 [22-27]. $E_{max}$ defined in accordance with the following guidelines: (1) a sharp peak neutron energy for quasi-monoenergetic neutrons such as p-Li sources, (2) the neutron energy with 1/100 of the neutron flux for that of the peak neutron energy in lethargy unit when the neutron spectrum has a clear broad peak, and (3) the neutron energy with 1/100 of the neutron flux for that at 20 MeV in lethargy unit otherwise. In the present case, since no clear peak is seen in the calculated source neutron spectra for the direction of 90° with respect to the proton beam direction, the value of Emax was determined to be 600 MeV by applying the guideline (3). The attenuation length is consistent with the previous data and it increases linearly with the maximum source neutron energy up to about 1 GeV [14].

Figure 5 shows neutron reaction rate distributions of the $^{209}$Bi (n, 4n) $^{206}$Bi reaction ($E_{th}$ ~ 22.6 MeV) along the proton beam line behind the iron and concrete shield as another typical experimental data. In this measurement, various activation detectors were set on the Concrete Cap (CC) and iron shield: Air Gap between concrete and iron (AG). After irradiation by the transmitted neutrons through the concrete and iron shields, radioactivity produced in each detector was also measured by HP-Ge detectors. A parameter of the Moyer Model, β, was also deduced from the measured reaction rate distributions by using the least square fitting in the angular region between 60 and 120 degrees which is applicable to the Moyer Model. The deduced β is about 4.0. It is larger than previously obtained value β = 2.3 [24]. A possible reason for this is energy dependence of β at energies above 100 GeV [14].

Transmitted neutron energy spectra through concrete and iron shields were measured by a multi-moderator spectrometer (Bonner spheres) using a pair of $BF_3$ proportional counters: $^{nat}BF_3$ (18% $^{10}$B) and $^{10}BF_3$ (96% $^{10}$B), of which gas pressure is 760 Torr. The detectors were covered by spherical moderators made of polyethylene (density of 0.928 g/cm$^3$) with the diameter of 81, 110, 150 and 230 mm. Because of the high neutron intensity at the top of the concrete, electronics for the detectors was operated as a current readout mode, in which input charge of 1 pC was converted to 1 pulse and time-dependent pulse numbers were counted by a multi-channel scaler. A higher Q-value of the $^{10}$B(n, α) reaction, 2.79 MeV, reduces the effect of contamination of the γ-rays, which was estimated to be less than 0.1 % by a measurement in a low-intensity neutron field of $^{241}$Am-Be with a pulse counting mode. Figure 6 shows the neutron energy spectrum obtained from the measurement by an unfolding method with the SANDII code [28] and response function calculated by the MCNPX code [29]. In the figure the neutron spectra calculated by the PHITS, MARS and MCNPX codes are compared with the measured spectrum. The calculations show an agreement with the measurement within a factor of two, although the PHITS and MARS calculations overestimate the measurement above a few MeV energy region while the MCNPX calculation underestimates the measurement overall [11].

The neutron reaction rates of In, Al, Bi and Cu measured by the activation method are compared with the reaction rates calculated by the neutron energy spectrum obtained by the Bonner sphere method, as shown in Fig. 7. The threshold energy of the $^{115}$In(n, n')$^{115m}$In reaction is about 0.6 MeV and the $^{209}$Bi(n, 7n)$^{203}$Bi reaction with threshold energy of about 45.3 MeV has the reaction probability up to about 100 MeV. It is shown that the results obtained by the Bonner sphere method are consistent with the results measured by the activation method in the energy region from 0.6 to about 100 MeV, and both experimental methods are reliable [11].

We are also performing a measurement of >100-MeV neutrons by using the Au activation method coupled with a low-background γ-ray counting system. As an indicator for the neutron flux, we determined the production rates of 8 spallation nuclides ($^{196}$Au, $^{188}$Pt, $^{189}$Ir, $^{185}$Os, $^{175}$Hf, $^{173}$Lu, $^{171}$Lu and $^{169}$Yb) having high threshold energies in the Au activation detector. From these results, we will deduce neutron spectra at neutron energy higher than 100 MeV.

### *3. 2. Measurement of particle flux, spectra and residual activity around shield at NuMI with 120-GeV protons*

At the NuMI experimental station, a 94-cm long graphite target irradiated with 120-GeV protons generates the secondary particles [30]. The beam power during the run was 150 kW, i.e. about a half of the NuMI beam capability. The secondary particles, mainly pions, are focused in the direction of a neutrino detector with two magnetic horns and lead to a 675-m-long decay pipe, in which pions decay into muons and neutrinos. The muons pass through a hadron absorber made of iron and concrete located downstream of the decay pipe, in which protons, neutrons and pions are removed from the beam axis. Figure 8 shows geometry of the experiment along the course of the flying muons. Behind the absorber, there are four caves at the flight distances of 0, 13.7, 33.5, and 67.1 m from the rock surface: “Alcove-1,” “Alcove-2,” “Alcove-3,” and “Alcove-4” [9].

Activation detectors were installed in every Alcove along the beam axis and were irradiated with muons and tertiary particles passing through the rock. In order to measure the radiation dose due to the muons and tertiary particles, OSL (Optical Stimulated Luminescence) dosimeter, TLDs (thermo-luminescence detectors, Panasonic 813 PQ), solid state nuclear track detectors, CR39, and ionization chamber (IC) were set along the beam axis, the passes from the Alcove to the bypass tunnel.[9]

Activation detectors were installed in every Alcove along the beam axis and were irradiated with muons and tertiary particles passing through the rock. In order to measure the radiation dose rates and reaction rates due to the muons and tertiary particles, OSL (Optical Stimulated Luminescence) dosimeter, TLDs (thermo-luminescence detectors, Panasonic 813 PQ), solid state nuclear track detectors, CR39, the Bonner sphere counters and ionization chamber (IC) for neutron and γ-ray survey were set along the beam axis, the passes to the Alcove and the bypass tunnel [9].

As a typical experimental data at NuMI, mass distribution of isotope production on a copper sample irradiated at Alcove-1 is shown in Fig. 9 [7]. The yields of 18 nuclides ($^{64}$Cu, $^{57}$Ni, $^{58}$Co, $^{57}$Co, $^{56}$Co, $^{55}$Co, $^{59}$Fe, $^{54}$Mn, $^{52}$Mn, $^{51}$Cr, $^{48}$V, $^{48}$Sc, $^{47}$Sc, $^{46}$Sc, $^{44m}$Sc, $^{43}$K, $^{42}$K, and $^{24}$Na) on copper samples were measured by the activation method. It is shown that the mass yield linearly increases with increasing the mass number between 40 and 60. The mass yield is larger than the linear slope in a lower-mass region and around copper. A reaction mechanism between muons and nuclei is still under detailed analysis of the mass distribution using the MARS15 code [7].

The average values of the yield ratios normalized to the value at Alcove-2 are plotted in Fig. 10 as a function of the depth from the rock surface. A variation of these ratios for the nuclides studied is within a factor of two. The yields from Alcove -2 show a gradual exponential decrease with the depth, while the yields decrease steeply between Alcove-1 and Alcove-2. Attenuation behavior at Alcove 2 to 4 is consistent with the calculation by the MARS15 code, and the profile of the yields is similar to the calculated attenuation of muons in the rock. An additional experiment is planned to see if neutrons and other particles can contribute to the measurements in Alcove-1 [7].

Dose distributions along the pass to the Alcove 2 are shown as a function of lateral distance from the beam axis in Fig. 11 [9]. In this figure, the points of OSL, TLDg and ionization chamber show measured dose from, mainly, muons, electrons and photons. The points of CR39 show fast neutron and CR39(B) and TLDnth thermal neutron. The experimental results indicates that the dose rate is dominated by muons, electrons and photons around the beam axis and the contribution of fast neutrons, which may be generated by high-energy muon interactions, increase with the distance from the beam line.

Dose rate distributions along the pass to the Alcove 2 are shown as a function of lateral distance from the beam axis in Fig. 11 [9]. OSL, TLD and ionization chamber have sensitivity to muons, electrons and photons. CR39 and CR39(B) can measure fast and thermal neutrons, respectively, and TLD can also detect thermal neutrons. It is shown that the dose rate is dominated by muons, electrons and photons around the beam axis and the contribution of fast neutrons, which may be generated by high-energy muon interactions, increase with the distance.

The experimental results are compared with the dose calculated by the MARS15 code in absolute value in the figure. Lines named as DET, DEM, DEE, DEN and DEG show the calculated dose of total, muon, electron, neutron and photon, respectively. The calculated total dose shows an excellent agreement with the dose rates by OSL and TLD around the beam axis and the distance larger than 7 m. A slight underestimation between 4 and 7 m can be caused by the difference in the calculation geometry. The calculation reveals that the total dose is mainly driven by muons around the beam axis. The electrons and photons are generated in the interactions of muons in the rock around the axis. The dose rate due to the electrons dominates the total dose rate at the distance from 2 m in the pass of the Alcove, because the electron generation has the angular distribution toward the lateral direction in the electromagnetic shower [9]. Both measurement and calculation show that contribution of fast neutrons drastically increase at the distance larger than 4 m. An explanation is that neutrons are generated by photo-nuclear reactions induced by photons generated by muons (bremsstrahlung and deep-inelastic nuclear reactions), of which the angular distribution is more isotropic compared to that of the electron generation.

## 4. Summary

Experimental results on particle fluxes, spectra and mass distributions were obtained at the Pbar and NuMI setups of FNAL. Shielding parameters - such as neutron attenuation length - were estimated and compared with the previous values. The results were analyzed by PHITS and MARS15, and the accuracy behind thick shield was confirmed within a factor of two. The MARS15 results are in a good agreement with experiments on dose distributions due to the particles generated by the high energy muons, with interaction mechanisms to be further analyzed. As a future plan, thick target yields and cross section measurements will be started at the Meson Test beam facility soon, based on the preliminary measurements. Finally, an irradiation field will be established for study on radiation damage.

### Acknowledgment

This work is supported by grant-aid of ministry of education (KAKENHI 19360432 and 21360473) in Japan. Fermilab is a U.S. Department of Energy Laboratory operated under Contract DE-AC02-07CH11359 by the Fermi Research Alliance, LLC.

### References

[1] H. Hirayama, et al., “Inter-comparison of the Medium–energy Neutron Attenuation in Iron and Concrete (5)”, Shielding Aspects of Accelerators, Targets and Irradiation Facilities (SATIF7), ITN, Lisbon, 17-18 May 2004, OECD/NEA Paris (2005), pp. 117-126.

[2] H. Hirayama, et al., “Inter-comparison of the Medium–energy Neutron Attenuation in Iron and Concrete (6)”, Shielding Aspects of Accelerators, Targets and Irradiation Facilities (SATIF8), Pohang, Korea, 22-24 June 2006, NEA/NSC/DOC(2010)6, pp. 237-249.

[3] N. Nakao, et al., “Measurement of neutron energy spectra behind shielding of a 120 GeV/c hadron beam facility”, NIM A562 (2006), 950-953.

[4] http://www.ganil.fr/eurisol/.

[5] http://projectx.fnal.gov/

[6] To be published in T. Sanami, et al., Proc. on Workshop on Radiation Detectors and Their Uses (2008).

[7] To be published in H. Matsumura, et al., Proc. on Workshop on Environmental Radioactivity (2008).

[8] H. Nakashima, et al, “Experimental Studies of Shielding and Irradiation Effects at High Energy Accelerator Facilities”, Nuclear Technology 168 (2009), 482-486.

[9] To be published in T. Sanami, et al., “Shielding Experiments at High Energy Accelerators of Fermilab (I) Dose rate around high intensity muon beam”, Proc. of the Fifth International Symposium on Radiation Safety and Detection Technology (2009).

[10] To be published in H. Yashima, et al., “Shielding Experiments at High Energy Accelerators of Fermilab (II): Spatial distribution measurement of reaction rate behind the shield and its application for Moyer model”, Proc. of the Fifth International Symposium on Radiation Safety and Detection Technology (2009).

[11] To be published in M. Hagiwara, et al., “Shielding Experiments at High Energy Accelerators of Fermilab (III): Neutron Spectrum Measurements in Intense Pulsed Neutron Fields of The 120-GeV Proton Facility Using A Current Bonner Sphere Technique”, Proc. of the Fifth International Symposium on Radiation Safety and Detection Technology (2009).

[12] To be published in N. Matsuda, et al., “Shielding Experiments at High Energy Accelerators of Fermilab (IV): Calculation Analyses”, Proc. of the Fifth International Symposium on Radiation Safety and Detection Technology (2009).

[13] D. J. Boehnlein, et al., “Studies of Muon-Induced Radioactivity at NuMI”, Proc. of the 11th International Workshop on Neutrino Factories, Superbeams and Beta Beams, the American Institute of Physics, Chicago, Illinois, U.S.A. (2010), 344.

[14] To be published in Y. Kasugai, et al., “Shielding Experiments under JASMIN Collaboration at FERMILAB (I) Overview of the Research Activities”, Proc. of International Conference on Nuclear Data for Science and Technology 2010 (2010).

[15] To be published in H. Yashima, et al., “Shielding Experiments under JASMIN Collaboration at FERMILAB (II) Radioactivity measurement induced by secondary particles from the anti-proton target”, Proc. of International Conference on Nuclear Data for Science and Technology 2010 (2010).

[16] To be published in H. Matsumura, et al., “Shielding Experiments under JASMIN Collaboration at FERMILAB (III) Measurement of High-Energy Neutrons Penetrating a Thick Iron Shield from the Antiproton Production Target by Au Activation Method”, Proc. of International Conference on Nuclear Data for Science and Technology 2010 (2010).

[17] To be published in H. Matsuda, et al., “Shielding Experiments under JASMIN Collaboration at FERMILAB (IV) Measurement and Analyses of High-Energy Neutrons Spectra in the Antiproton Production Target Station”, Proc. of International Conference on Nuclear Data for Science and Technology 2010 (2010).

[18] K. Niita, et al., “PHITS—a particle and heavy ion transport code system”, Radiat. Meas. 41(2006), 1080-1090.

[19] N. V. Mokhov, FERMILAB-FN-628 (1995), N.V. Mokhov, S.I. Striganov, Fermilab-Conf-07-008-AD (2007), Hadronic Shower Simulation Workshop AIP Proceedings 896.

[20] To be published in T. Sanami, et al., “Neutron energy spectrum from 120 GeV protons on a thick copper target (JASMIN in M-test)”, Shielding Aspects of Accelerators, Targets and Irradiation Facilities (SATIF10), CERN, Swiss, 2-4 June (2010).

[21] B. J. Moyer, “Method of Calculation of the Shielding Enclosure for the Berkeley Bevatron,” Proc. 1st Int. Conf. Shielding around High Energy Accelerators, Presses Universitaires de France, Paris (1962), 65.

[22] M. Sasaki, et al., “Measurements of High energy Neutrons Penetrated Through Iron Shields Using Self-TOF Detector and NE213 Organic Liquid Scintillator”, NIM B196 (2007), 113-124.

[23] S. Ban et al., “Measurement of transverse attenuation lengths for paraffin, heavy concrete and iron around an external target for 12 GeV protons”, NIM 174 (1980), 271-276.

[24] G. R. Stevenson, et al., “Determination of transverse shielding for proton accelerators using the Moyer model”, Health Physics 43 (1982), 13-29.

[25] T. Nunomiya, et al., “Measurement of deep penetration of neutrons produced by 800-MeV proton beam through concrete and iron at ISIS”, NIM B 179(2001), 89-102.

[26] N. Nakao, et al., “Transmission Through Shields of Quasi-Monoenergetic Neutrons Generated by 43 and 68-MeV Protons - I: Concrete Shielding Experiment and Calculation for Practical Application”, NSE 124 (1996), 228-242.

[27] T. Ishikawa, et al., “Neutron Penetration Through Iron and Concrete Shields with the Use of 22.0- and 32.5-MeV Quasi-Monoenergetic Sources”, NSE 116 (1994), 278.

[28] W. McElroy, S. Breg and G. Gigas, "Neutron-Flux Spectral Determination by Foil Activation," NSE 27(1967), 533-541.

[29] J. S. Hendricks, et al., LA-UR-08-2216 (2008); http://mcnpx.lnal.gov/

[30] K. Anderson, et al., “NuMI facility technical design report, Technical report”, FERMILAB-AB-DESIGN-1998-01, Fermilab (1998).

Table 1: Size and purity of activation detectors used in activation method

| Activation Foil | Purity (%) | Size (mm) |
|---|---|---|
| Carbon (C) | 99.99 | $\phi$ 80x10 |
| Indium (In) | 99.99 | $\phi$ 80x10 |
| Aluminum (Al) | 99.999 | $\phi$ 80x10 |
| Copper (Cu) | 99.994 | $\phi$ 80x10 |
| Bismuth (Bi) | 99.5 | $\phi$ 80x10 |

**Figure 1: Comparison of the neutron attenuation length of iron [1]**

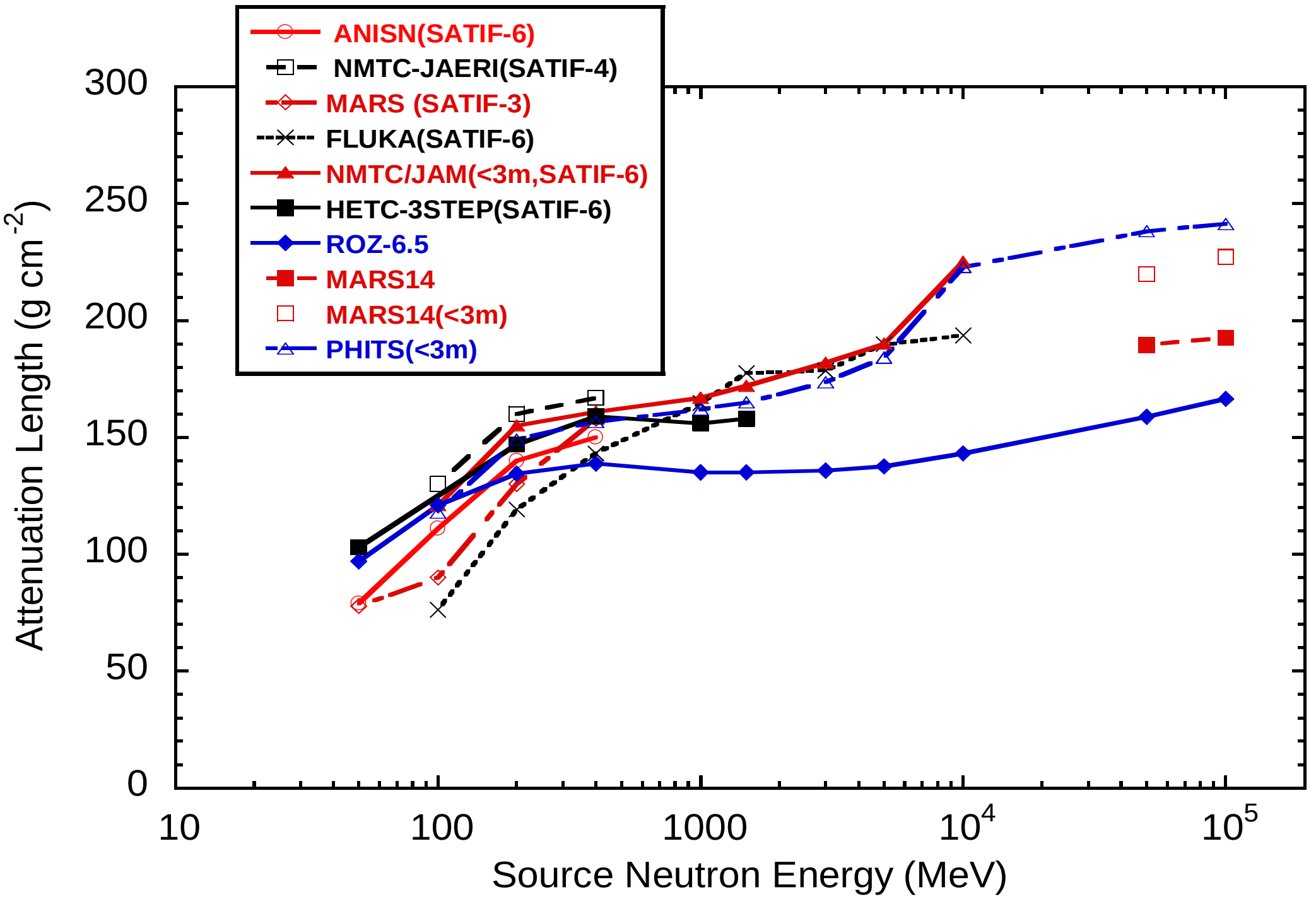

**Figure 2: Cross sectional view of the Pbar target station [8]**

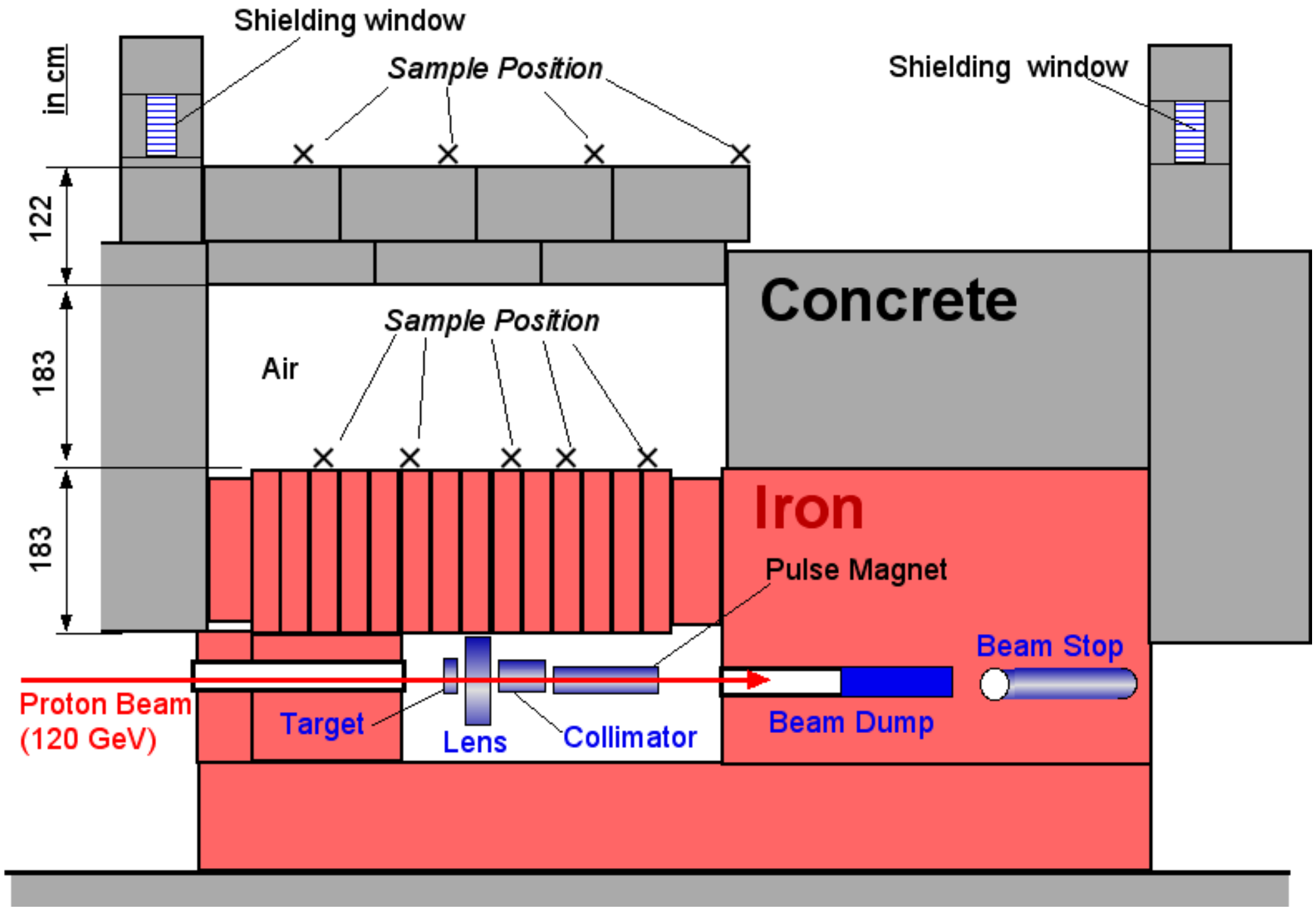

**Figure 3: Reaction rate distributions in iron shield**

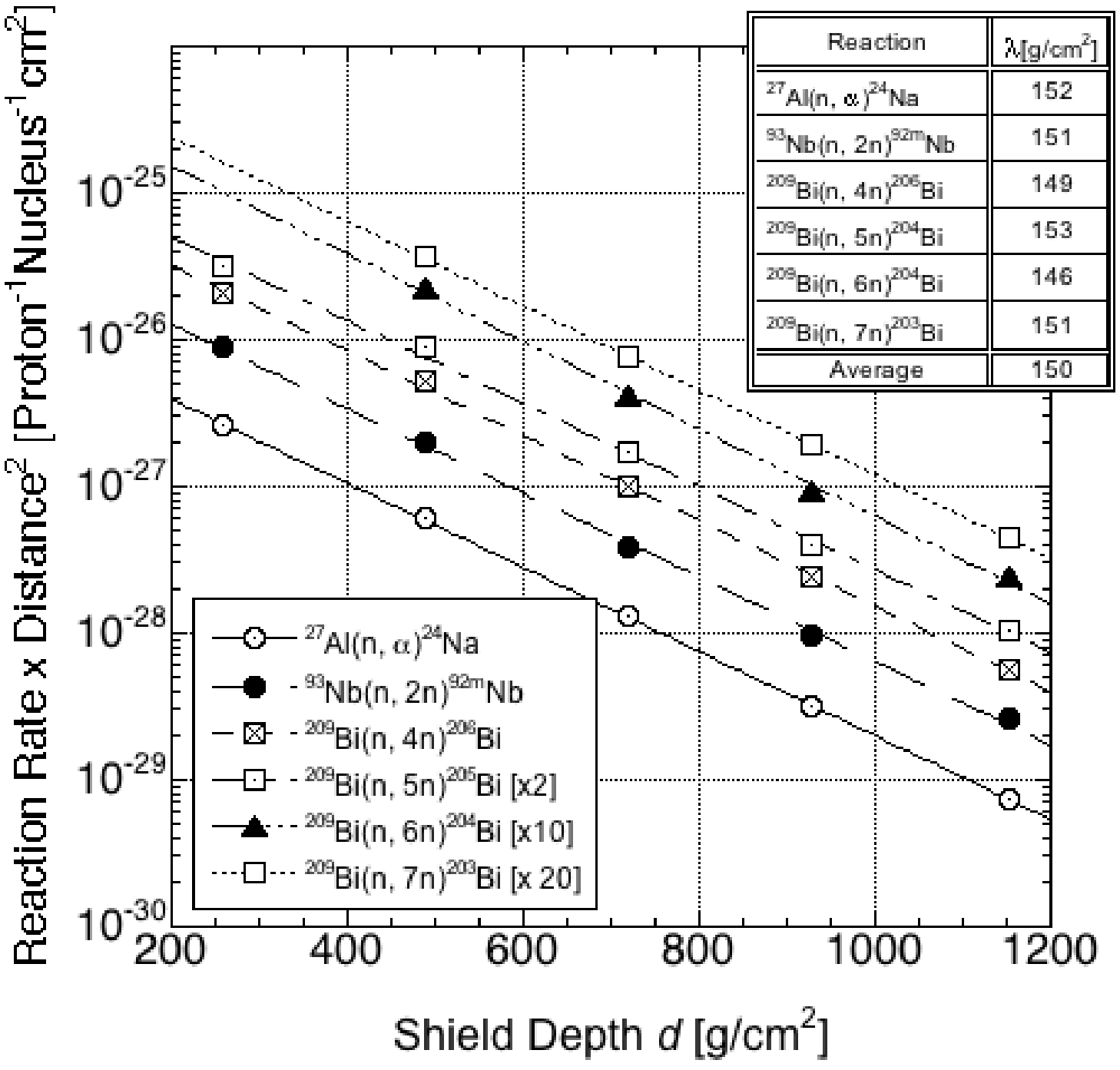

**Figure 4: Comparison of attenuation length as function of maximum source neutron energy**

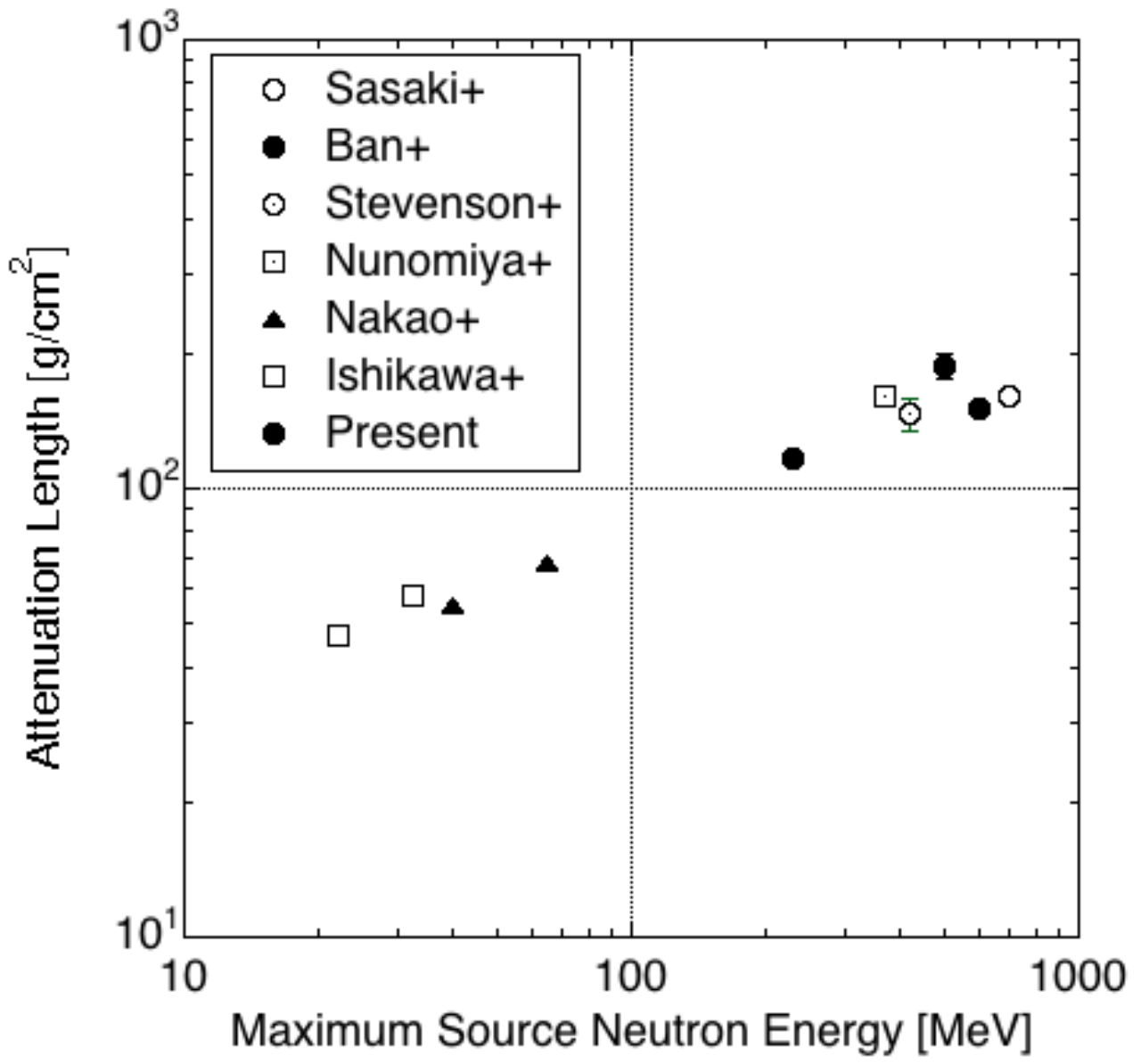

**Figure 5: Reaction rate distribution of $^{209}$Bi (n, 4n)$^{206}$Bi. Open and closed circles show experimental data at the surface of iron shield (AG) and concrete (CC). The fitting curve by the Moyer model are also shown with the solid lines [14].**

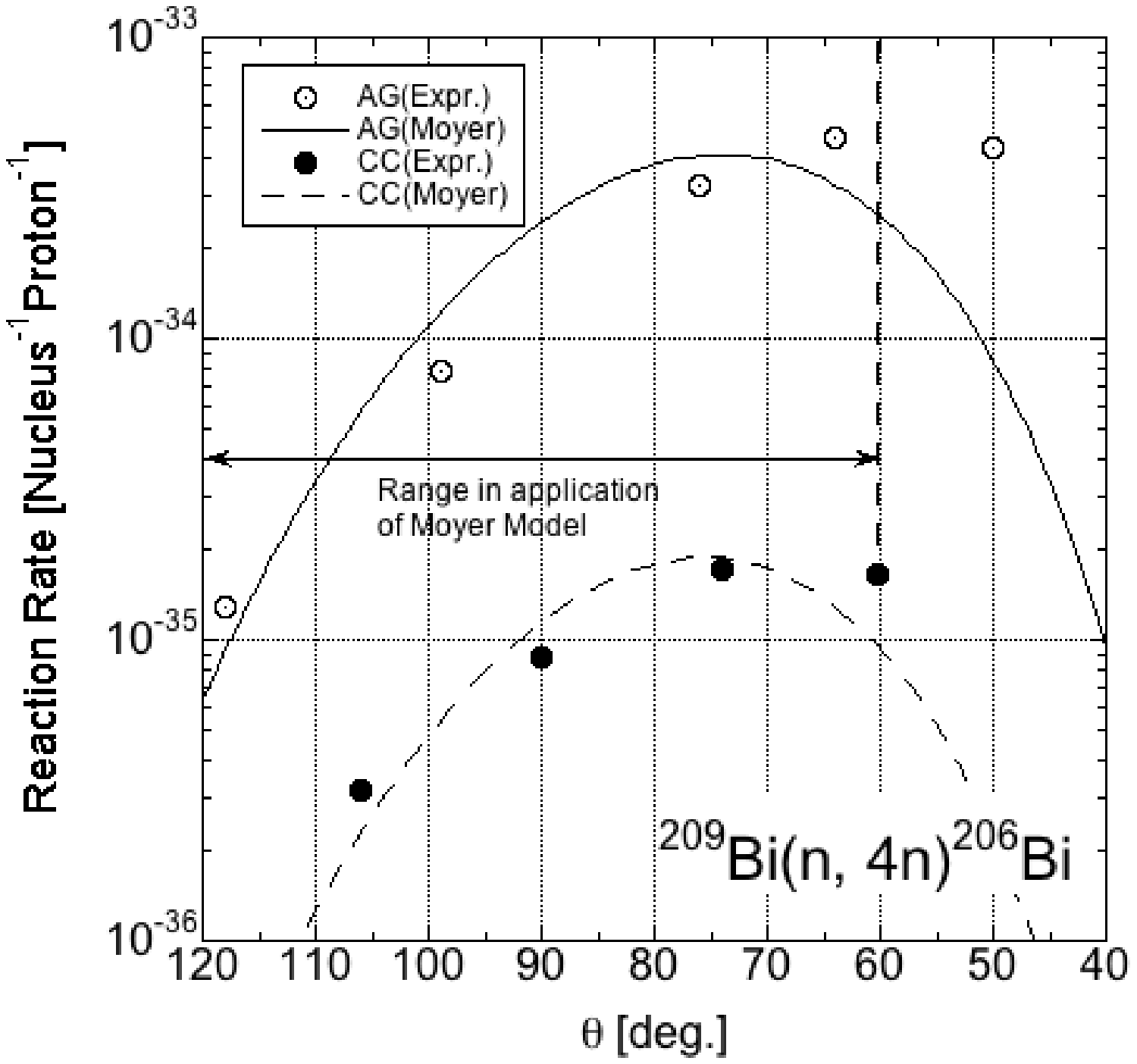

**Figure 6: Neutron energy spectra above the Pbar target station measured by the Bonner spheres and calculated by the PHITS, MARS15 and MCNPX codes.**

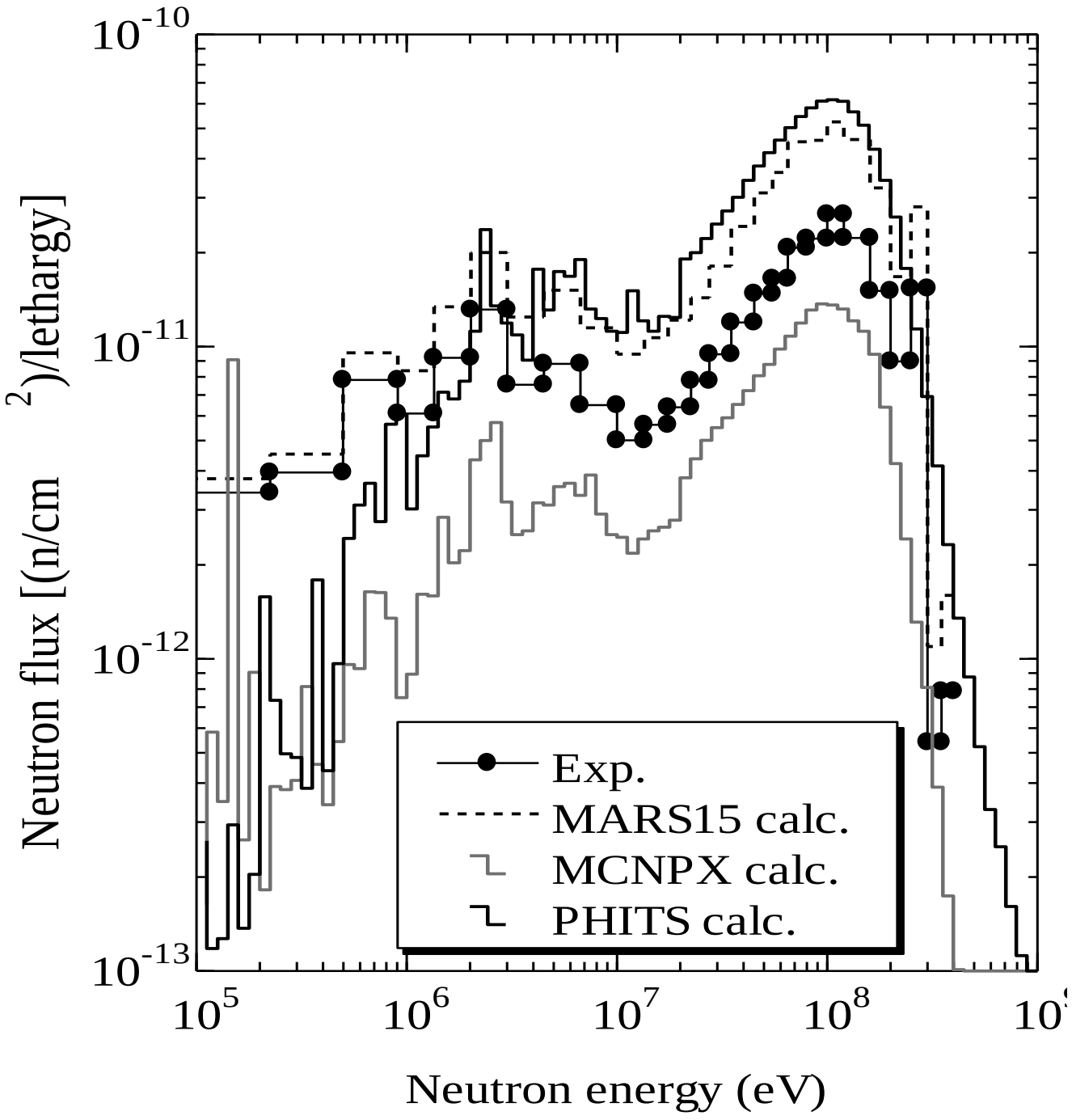

**Figure 7: Comparison of reaction rates measured by the Bonner spheres and activation method [11].**

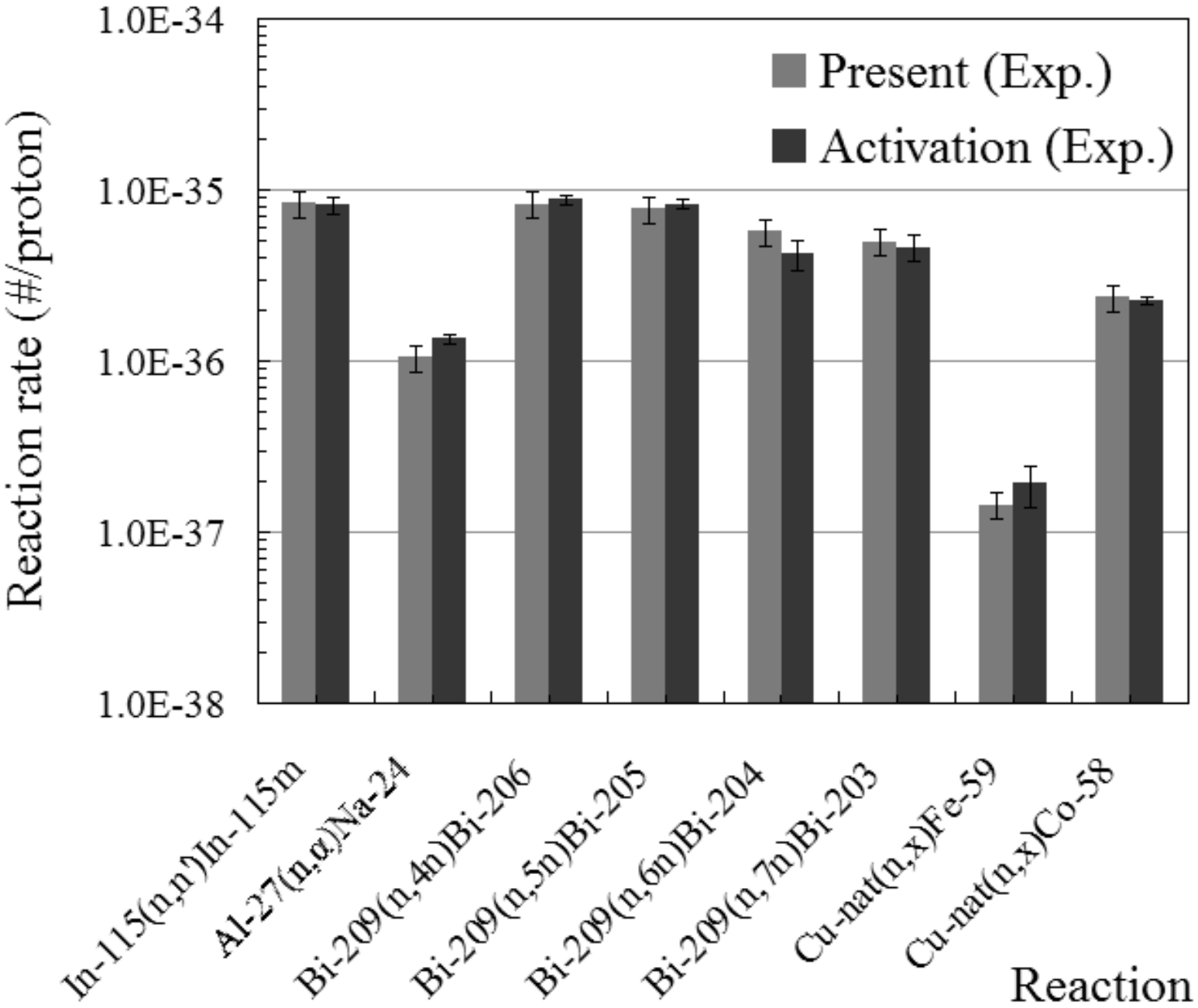

**Figure 8: Schematic view of downstream of NuMI and location of dosimeters and detectors [9].**

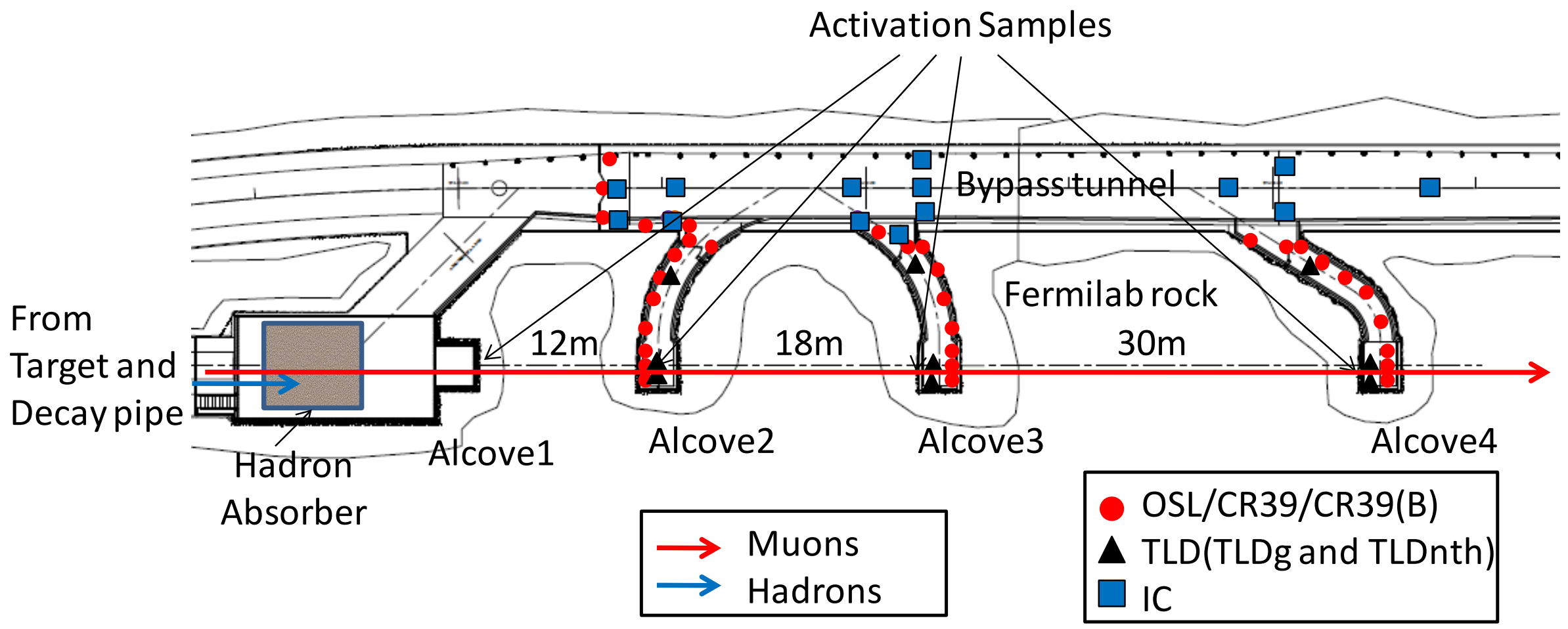

**Figure 9: Mass yield distribution on Cu at Alcove-1 [7].**

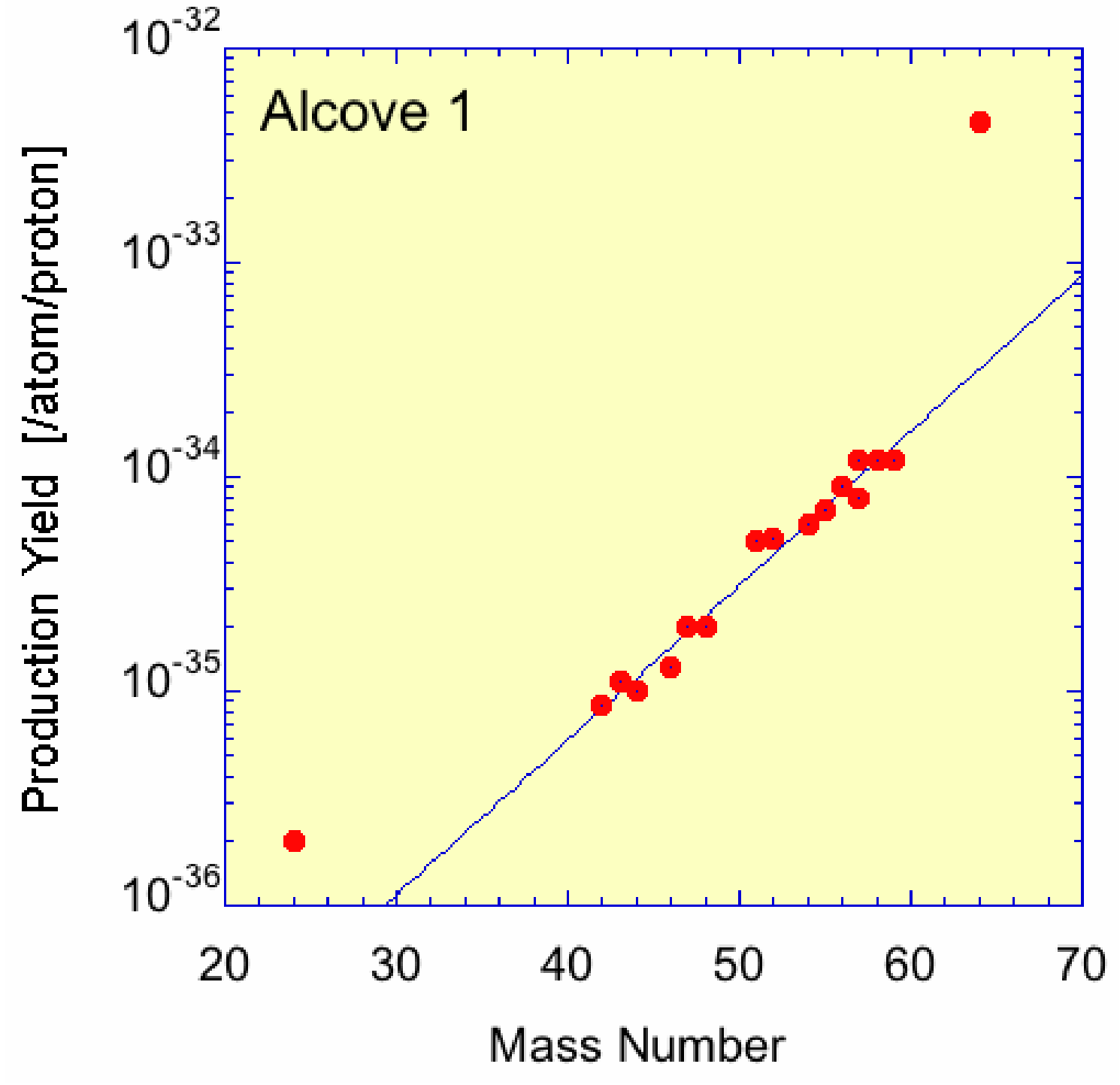

**Figure 10: Attenuation profile of the mass yields in rock, which is normalized by the yield at Alcove-2. The solid line is drawn to guide the eye [7].**

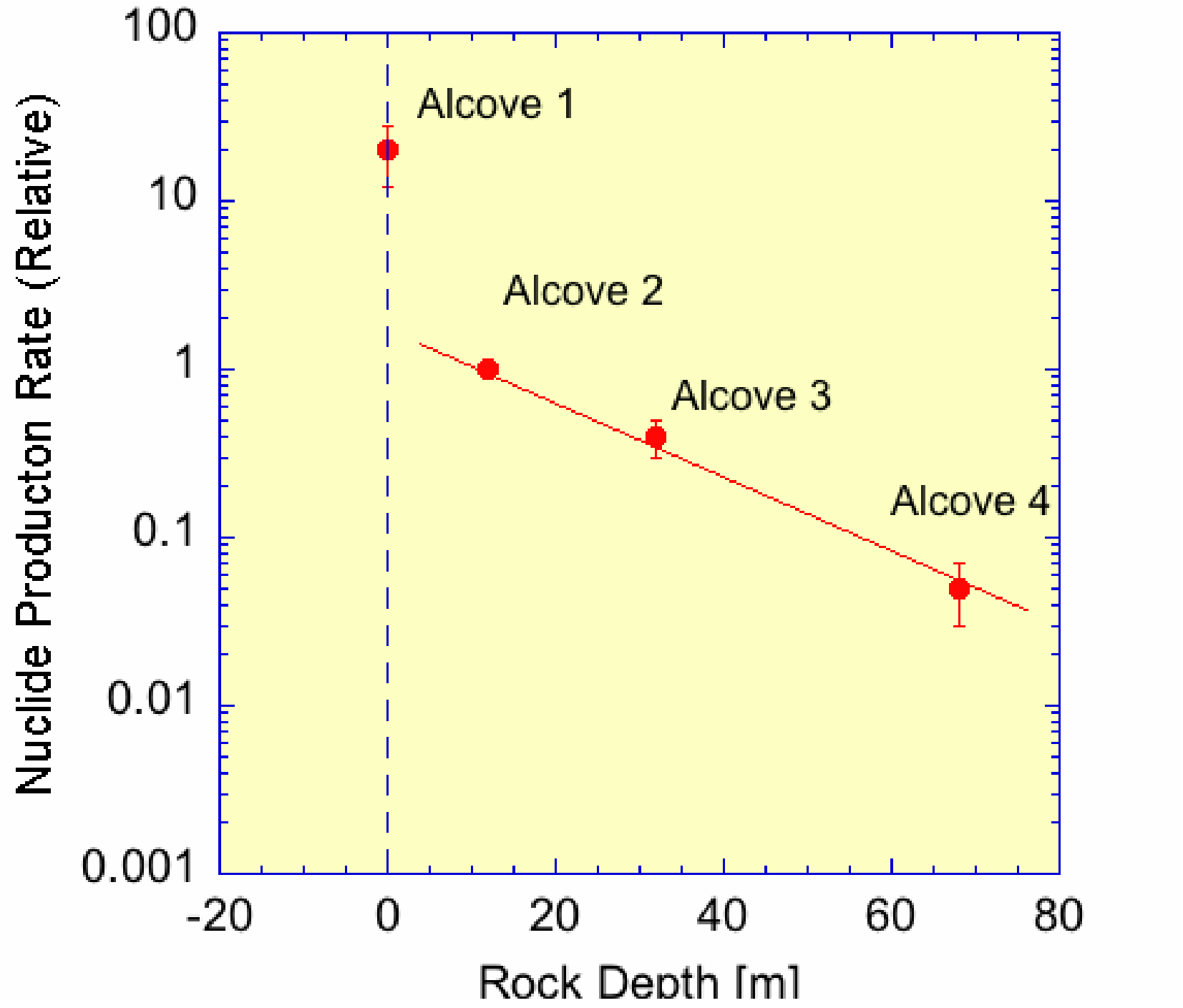

**Figure 11: Dose rate distributions measured by various dosimeters along Alcove-2, which is compared with those of total, muon, photon, electron and neutron calculated by the MARS code [9].**

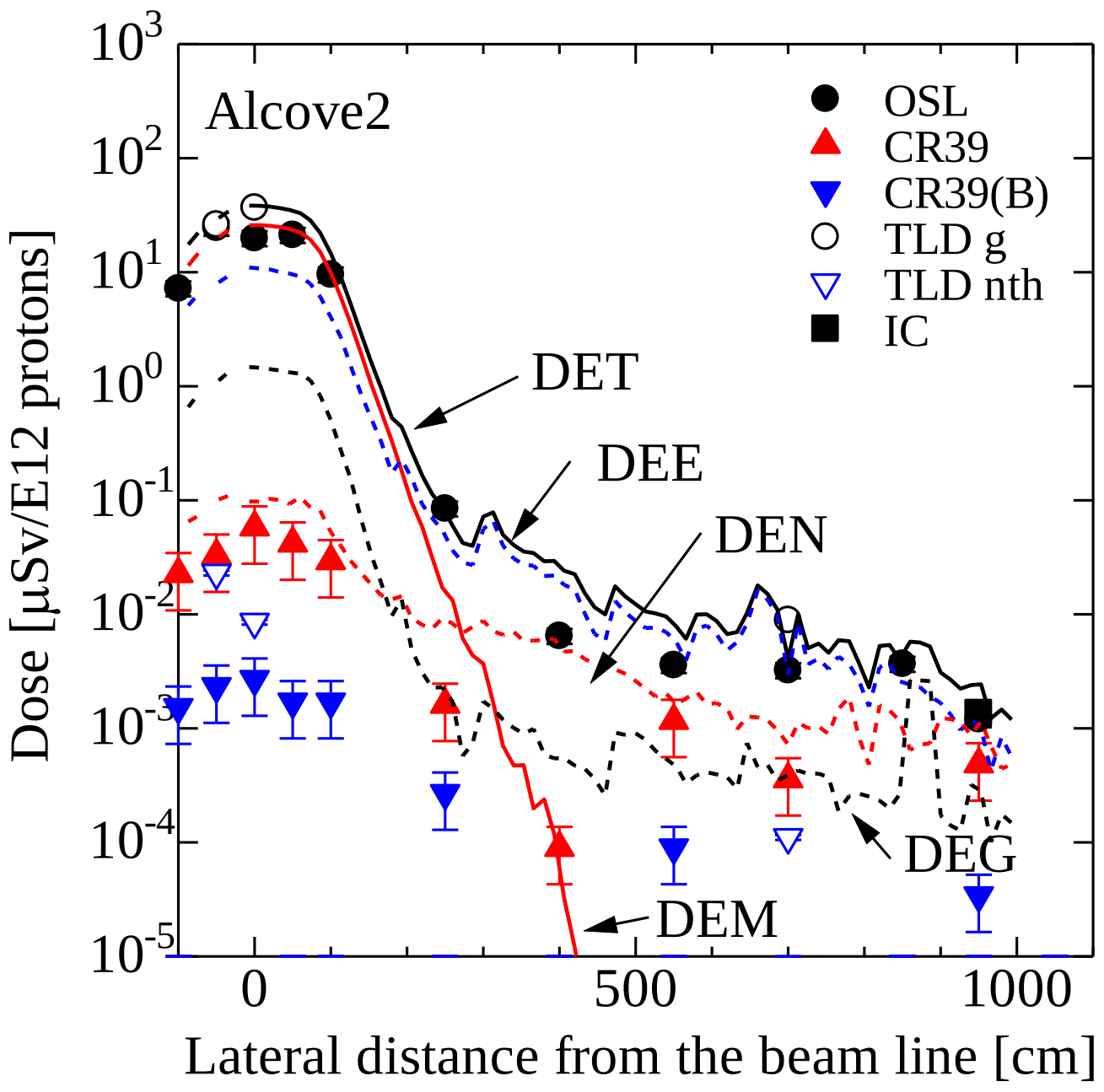